\documentstyle[11pt,aaspp4,flushrt]{article}
\catcode`\@=11 
\def\@versim#1#2{\vcenter{\offinterlineskip
        \ialign{$\m@th#1\hfil##\hfil$\crcr#2\crcr\sim\crcr } }}
\newcommand{\beq}{\begin{equation}}
\newcommand{\eeq}{\end{equation}}

\def\lsim{\mathrel{\mathpalette\@versim<}}
\def\gsim{\mathrel{\mathpalette\@versim>}}
\def\H{\tilde H}

\begin{document}
\title{Angular Momentum Transport in Particle and Fluid Disks}
\author{Eliot Quataert\footnote{Chandra Fellow}$^,$\footnote{Institute for Advanced Study, School of Natural Sciences, Einstein Drive, 
Princeton, NJ 08540; eliot@ias.edu} and Eugene I. Chiang\footnote{Theoretical Astrophysics, Caltech 130-33, Pasadena, CA 91125; echiang@tapir.caltech.edu}}
\centerline{}
\centerline{\it To appear in the Astrophysical Journal:  Nov. 20, 2000}

\begin{abstract}

We examine the angular momentum transport properties of disks composed
of macroscopic particles whose velocity dispersions are externally
enhanced (``stirred'').  Our simple Boltzmann equation model serves as
an analogy for unmagnetized fluid disks in which turbulence may be
driven by thermal convection.  We show that interparticle collisions
in particle disks play the same role as fluctuating pressure forces
and viscous dissipation in turbulent disks: both transfer energy in
random motions associated with one direction to those associated with
another, and convert kinetic energy into heat.  The direction of
angular momentum transport in stirred particle and fluid disks is
determined by the direction of external stirring and by the properties
of the collision term in the Boltzmann equation (or its analogue in
the fluid problem).  In particular, our model problem yields inward
transport for vertically or radially stirred disks, provided
collisions are suitably inelastic; the transport is outwards in the
elastic limit. Numerical simulations of hydrodynamic turbulence driven
by thermal convection find inward transport; this requires that
fluctuating pressure forces do little to no work, and is analogous to
an externally stirred particle disk in which collisions are highly
inelastic.

\end{abstract}

\section{Introduction}

When turbulence in unmagnetized Keplerian disks is driven by thermal
convection, angular momentum appears to be transported inwards (Ryu \&
Goodman 1992; Kley, Papaloizou, \& Lin 1993; Stone \& Balbus 1996;
Cabot 1996).  If generically true, this implies that convection cannot
provide the ``enhanced'' angular momentum transport (relative to
collisional viscosity) needed to account for observations of accreting
systems in astrophysics.

Despite appearances, inward transport does not contradict the fact
that an isolated Keplerian disk relaxes to its minimum energy state by
transporting angular momentum outwards (Lynden-Bell \& Pringle 1974).
The reason is that simulated convective disks are not completely
isolated. They tap a source of free energy other than their intrinsic
differential rotation: an externally imposed, superadiabatic entropy
gradient.

A more mechanistic understanding of the behavior of turbulent fluid
disks can be obtained by considering disks composed of discrete
macroscopic particles.  Transport properties and stability criteria
obtained under one picture often carry over to the other, to within
dimensionless factors of order unity. The fluid/particle dualism has
been employed with considerable success in studies of planetary rings
(see, e.g., Goldreich \& Tremaine 1982) and galactic structure (see,
e.g., Chapter 6 of Binney \& Tremaine 1987).

Indeed, one-to-one correspondences can be established between terms in
the Reynolds stress tensor equation describing turbulent fluid disks,
and terms in the Boltzmann stress tensor equation describing particle
disks.  This paper explores the transport properties of the latter
system, with the aim of shedding light on the former.  We demonstrate
how simple particle disk models can reproduce many of the transport
phenomena found in numerical simulations of vertically or radially
convective fluid disks.  The particle viewpoint also provides insight
into the stability properties of isolated, unmagnetized, fluid disks.
An analysis similar to ours was carried out by Kumar, Narayan, \& Loeb
(1995); this is discussed in \S5.

In \S2, we write down the Boltzmann stress tensor equation governing
particle disks and discuss its intimate relationship with the Reynolds
stress tensor equation governing hydrodynamic turbulence in a fluid
disk.  In \S3, we prove a general theorem that relates the direction
of angular momentum transport to the behavior of the velocity
dispersion tensor in the presence of external driving.  Section 4 is
the most important of this paper; there we consider a simple model for
the collision term in the Boltzmann equation that allows one to easily
solve for the stress tensor of the disk and, we believe, capture much
of the relevant physics regarding both stability and the direction of
transport.  In \S5, we summarize and discuss our results.

\section{Stress Tensor Equations for Particle Disks}

The second moments of the Boltzmann equation describe the evolution of
the viscous stress tensor.  For an axisymmetric disk, in cylindrical
coordinates, these read (see, e.g., Borderies, Goldreich, \& Tremaine
1983):

\beq \partial_t p_{rr} - 4 \Omega p_{r\phi} = \left( \partial_t p_{rr}
\right)_c + \dot H_{rr} \label{rr} \eeq
\beq \partial_t p_{\phi \phi} +
{\kappa^2 \over \Omega} p_{r\phi} = \left( \partial_t p_{\phi \phi}
\right)_c + \dot H_{\phi \phi} \eeq \beq \partial_t p_{zz} = \left(
\partial_t p_{zz} \right)_c + \dot H_{zz}
\label{zz} \eeq
\beq \partial_t p_{r\phi} +
{\kappa^2 \over 2 \Omega} p_{rr} - 2 \Omega p_{\phi \phi} = \left(
\partial_t p_{r\phi} \right)_c \, , \label{rp} \eeq

\noindent where we have neglected third moments.  In equations
(\ref{rr})--(\ref{rp}), $\Omega$ is the angular frequency, $\kappa^2 =
r^{-3} d(\Omega r^2)^2/dr$ is the square of the epicyclic frequency,
and $p_{i j}$ are the components of the stress tensor.  The terms
$\left( \partial_t p_{i j} \right)_c$ describe changes in velocity
dispersions due to interparticle collisions. Collisions perform three
tasks: they (1) extract energy from the mean shear into random
motions, (2) transfer energy in random motions associated with one
direction to those associated with another, and (3) convert kinetic
energy into heat.  The terms $\dot H_{i j}$ are inserted to simulate
external stirring of the disk, with $\dot H_{r \phi} = 0$ so as not to
explicitly induce angular momentum transport.

\subsection{Boltzmann vs. Reynolds}

Since the fluid equations are derived from the Boltzmann equation it
is not surprising that hydrodynamic turbulence in fluid disks
satisfies energy conservation equations very similar to equations
(\ref{rr})--(\ref{rp}).  The Reynolds stress equation can be obtained
by taking the Navier-Stokes equation, decomposing the velocity and
pressure into mean and fluctuating components, and multiplying the
entire equation by the velocity.  This yields (e.g., Speziale 1991;
Kato \& Yoshizawa 1997): \beq \partial_t T_{ij} = -T_{ik} \partial_k
U_j - T_{jk} \partial_k U_i + \pi_{ij} - \epsilon_{ij} - \partial_k
C_{ijk}, \label{reynolds} \eeq where $U$ is the mean velocity of the
fluid, $u$ is the fluctuating velocity, $T_{ij} = \langle u_i u_j
\rangle$ is the stress tensor, and $\langle \ \rangle$ denotes an
ensemble average. The quantity $C_{ijk}$ is a sum of turbulent and
viscous fluxes, \beq \pi_{ij} = - \langle u_i \partial_j p + u_j
\partial_i p \rangle \eeq and \beq \epsilon_{ij} = 2 \nu \langle
\partial_k u_i \partial_k u_j \rangle, \eeq where $p$ is
the fluctuating gas pressure and $\nu$ is the kinematic viscosity.
The first two terms on the right hand side of equation
(\ref{reynolds}) describe how turbulence extracts energy from the
shear in the mean flow.  The terms $\pi_{ij}$ are known as the
``pressure-strain'' correlation in the fluid dynamics
literature;\footnote{To be precise, our $\pi_{ij}$ is related to the
usual pressure-strain correlation tensor by a total divergence, which
we have absorbed into $C_{ijk}$.} the diagonal terms of $\pi_{ij}$ are
simply the work done by fluctuating pressure forces associated with
the turbulence; among their other tasks, these terms attempt to
enforce isotropy of the turbulence by transferring energy from
velocity fluctuations associated with one direction to those
associated with another.  The terms $\epsilon_{ij}$ represent the
viscous dissipation of energy on small scales (note that
$\epsilon_{ii}$ is positive definite).

It is straightforward to evaluate the above expression for $\partial_t
T_{ij}$ in cylindrical coordinates. We neglect the flux
divergence\footnote{This assumes ``homogeneous'' turbulence, i.e.,
that the flux varies on a length scale $\sim r$, much greater than
the vertical scale height of the disk.} to obtain \beq \partial_t
T_{rr} - 4 \Omega T_{r\phi} = \pi_{rr} - \epsilon_{rr}
\label{rrh} \eeq \beq \partial_t T_{\phi \phi} + {\kappa^2 \over
\Omega} T_{r\phi} = \pi_{\phi \phi} - \epsilon_{\phi \phi} \eeq \beq
\partial_t T_{zz} = \pi_{zz} - \epsilon_{zz} \label{zzh} \eeq \beq
\partial_t T_{r\phi} + {\kappa^2 \over 2 \Omega} T_{rr} - 2 \Omega
T_{\phi \phi} = \pi_{r \phi} - \epsilon_{r \phi}
\label{rph} \, . \eeq 
Equations (\ref{rrh})--(\ref{zzh}) are identical to
the energy equations given by, e.g., Balbus \& Hawley (1998; see their
equations 58--60).

Equations (\ref{rr})--(\ref{rp}) for the evolution of the stress
tensor in a particle disk are very similar to equations
(\ref{rrh})--(\ref{rph}) for the evolution of the stress tensor in a
fluid disk.  In particular, the rotational profile of the disk enters
identically in the two cases; both $\Omega$ and $\kappa^2$ couple to
the stress tensor components in exactly the same way for particle and
fluid disks.  Moreover, the ``source'' and ``sink'' terms in the
energy equations are analogous; the collision term, $\left( \partial_t
p_{i j} \right)_c$, in the Boltzmann equation plays the same role as
$\pi_{ij} - \epsilon_{ij}$ in the Reynolds stress equations.

This comparison indicates that the dynamical role of epicycles is the
same in both fluid and particle disks, and that any difference in the
angular momentum transport properties of the two disks can only be due
to differences in the sources and sinks in the energy equation. We
demonstrate this formally in the next section by proving a theorem
that relates the direction of angular momentum transport to the
properties of the collisional and stirring terms.  This theorem is very
useful for interpreting the model problem of \S4.

\section{The ``Long Axis Must Get Longer'' Theorem}

{\it Theorem:} For thin axisymmetric disks in which $d \Omega/dr < 0$,
steady inward transport of angular momentum requires that the net
tendency of collisions and external stirring be to further lengthen the
principal equatorial long axis of the velocity ellipsoid and to
further shorten the principal equatorial short axis.  This theorem
applies to fluid disks as well as to particle disks so long as the
interparticle collision term, $(\partial_t p_{ij})_c$, is replaced by
$\pi_{ij} - \epsilon_{ij}$.

{\it Proof:} We transform equations (\ref{rr})--(\ref{rp})
to the principal axis basis of the velocity
ellipsoid.  Let $p_{11}$ and $p_{22}$ be the greater and lesser of the
two components of the stress tensor, respectively, in the equatorial
plane:
\begin{eqnarray}
p_{11} & = & [\, p_{rr} + p_{\phi \phi} + \sqrt{ (p_{rr} - p_{\phi \phi} )^2 + 4p_{r\phi}^2} \,] \, / \, 2  \\
p_{22} & = & [\, p_{rr} + p_{\phi \phi} - \sqrt{ (p_{rr} - p_{\phi \phi} )^2 + 4p_{r\phi}^2} \,] \, / \, 2 \, , 
\end{eqnarray}

\noindent where we shall always take the positive square root. By symmetry
about the midplane, $p_{33} = p_{zz}$. The orientation of the velocity
ellipsoid is given by $\delta$,
the angle between the longer equatorial axis and the radial direction:
\begin{eqnarray}
\tan \delta & \equiv & \frac{p_{r\phi}}{p_{11} - p_{\phi \phi}} \nonumber \\
 & = & \frac{2p_{r\phi}}{(p_{rr} - p_{\phi \phi}) +
\sqrt{(p_{rr} - p_{\phi \phi})^2 + 4p_{r\phi}^2}} \, ,\label{delta}
\end{eqnarray}

\noindent valid for $|\delta| < \pi/2$ and nonisotropic
$p_{ij}$.\footnote{If $\delta = \pm \pi/2$ or $p_{ij}$ is isotropic,
$p_{r\phi} = 0$ and there is no transport.}  Since the denominator in
equation (\ref{delta}) is positive, \beq \mbox{sgn} \, p_{r\phi} =
\mbox{sgn} \, \delta \, . \eeq
\noindent Outward (inward) transport corresponds to positive
(negative) $p_{r \phi}$, which in turn corresponds to positive
(negative) $\delta$.

In the principal axis basis, the viscous stress equations (\ref{rr})--(\ref{rp})
read (cf. Goldreich \& Tremaine 1978):
\beq \partial_t p_{11} - S\Omega \, (\sin 2\delta) \, p_{11} \, = \, \left(\partial_tp_{11}\right)_c + \dot H_{11} \label{t1}\eeq
\beq \partial_t p_{22} + S\Omega \, (\sin 2\delta) \, p_{22} \, = \, \left(\partial_tp_{22}\right)_c + \dot H_{22} \label{t2}\eeq
\beq \partial_t p_{33} \, = \, \left(\partial_t p_{33}\right)_c  + \dot H_{33} \label{t3}\eeq
\beq (p_{11} - p_{22}) \, \partial_t\delta - (2-S\sin^2\delta) \, \Omega \, p_{22} + (2-S\cos^2\delta) \, \Omega \, p_{11} \, = \, (p_{11}-p_{22}) \, (\partial_t\delta)_c + \dot H_{12}  \, , \label{tdelta}\eeq
\noindent where $S = - d \ln\Omega/d \ln r$.

Equations (\ref{t1}) and (\ref{t2}) state that for $S > 0$, the steady
($\partial_t = 0$) inward transport ($\delta < 0$) of angular momentum
demands that the net tendency of collisions and external stirring be
to further lengthen the principal long axis of the velocity ellipsoid
and to further shorten the principal short axis.

Collisions between hard, imperfectly elastic spheres tend to
isotropize the velocity ellipsoid: $(\partial_t p_{11})_c < 0 <
(\partial_t p_{22})_c$. Thus, in the absence of stirring, transport is
normally outwards in axisymmetric particle disks.

\section{Angular Momentum Transport in Externally Stirred Disks}

In this section we solve for the steady state structure of a particle
disk using a simple model for the effects of interparticle collisions
and external stirring. We also discuss the fluid analogue of our
particle disk model in some detail.

\subsection{The Particle Scattering Function}

Following, e.g., Cook \& Franklin (1964), Shu \& Stewart (1985), and
Narayan, Loeb, \& Kumar (1994) we take the scattering function in the
Boltzmann equation to be given by the Krook approximation: \beq
(\partial_t f)_c = {f_0 - f \over \tau},\eeq where $\tau$ is the
collisional mean free time, $f$ is the particle distribution function,
and $f_0$ is the ``equilibrium'' or ``post-scattering'' distribution
function.  The corresponding collision term in the viscous stress
equations is $\left( \partial_t p_{i j} \right)_c = (p^0_{i j} - p_{i
j})/\tau$.  The physical idea behind the Krook model is that when
particles collide they are removed from phase space at a rate
$-f/\tau$ and are returned to phase space with the new distribution
function $f_0$.

We assume that collisions do not generate off diagonal stresses so
that $p^0_{ij} = 0$ if $i \ne j$.  We account for the fact that
particle collisions are inelastic, converting some fraction of the
kinetic energy of the colliding particles into heat.  With this model,
the relevant second moments of $f_0$ are given by (see Kumar et
al. 1995) \beq p^0_{rr} = (1 - \xi_r) { \sigma^2 \over 3}, \ \ \
p^0_{\phi \phi} = (1-\xi_\phi) { \sigma^2 \over 3}, \ \ {\rm and} \ \
p^0_{zz} = (1 - \xi_z) { \sigma^2 \over 3},
\label{scat} \eeq where $\sigma$ is the velocity dispersion ($\sigma^2
= p_{rr} + p_{\phi \phi} + p_{z z}$) and the parameters $\xi_i \
\epsilon \ [0,1]$ measure the inelasticity of the collisions; $\xi
\sim 0$ is elastic, while $\xi \sim 1$ is highly inelastic.

Equation (\ref{scat}) allows the inelasticity of colliding particles
to be anisotropic, i.e., to be different for the $r$, $\phi$, and $z$
directions.  Such an anisotropic scattering function is unrealistic
for real particle disks in nature (e.g., planetary rings); collisions
between hard spheres at the microscopic level cannot be expected to
respect the cylindrical coordinate system.  Our strategy is to first
solve for the structure of ``real'' externally stirred particle disks,
in which the $\xi_i$'s are equal.  This captures much of the physics
in which we are interested.  We will show, however, that the
assumption of equal $\xi_i$'s does not furnish a fully adequate
model of fluid disks and that the introduction of an ``anisotropic
inelasticity'' can remedy this defect.

\subsection{The Fluid Analogue of the Scattering Function}
\label{flanalog}

Our particle scattering function has a mathematically rigorous
analogue in fluid turbulence.  Namely, it is equivalent to a
particular closure scheme for the {\it a priori} unknown turbulent
quantities $\epsilon_{ij}$ and $\pi_{ij}$: \beq \epsilon_{ij} =
{T_{ij} \over \tau} \ \ \ {\rm and} \ \ \ \pi_{ij} = {(1 - \xi_i)
\over 3 \tau} \sigma^2 \delta_{ij}. \label{turb}\eeq Interpreted as
such, $\tau$ is roughly the timescale on which free turbulence would
decay (the eddy turnover time) and the inelasticity parameter,
$\xi_i$, measures the efficiency with which fluctuating pressure
forces perform work in the $i-$th direction.  The closure model
proposed in equation (\ref{turb}) differs from conventional
second-order closure models of the Reynolds stress equation (e.g.,
Speziale 1991; see Kato \& Yoshizawa 1997 for an application to
Keplerian disks).  We will show, however, that this simple model
provides considerable insight into the angular momentum transport
properties of accretion disks.

In contrast to $\tau$, it is not obvious what values to assign $\xi_i$
in hydrodynamic turbulence. Although in one sense turbulence is
``inelastic,'' with energy cascading to small scales in the absence of
source terms, the physical meaning of $\xi_i$ defined by equation
(\ref{turb}) is more precise: fluid turbulence is ``inelastic'' if and
only if fluctuating pressure forces are inefficient at doing work.
Throughout the remainder of this paper it is worth keeping in mind
that numerical simulations suggest that the highly ``inelastic'' limit
is appropriate for Keplerian disks.  Note also that, unlike in the
particle disk case, we would not expect the $\xi_i$'s to be identical
in the fluid case.  The fluctuating pressure forces are macroscopic in
origin and thus can plausibly be different for the $r$, $\phi$, and
$z$ directions.

\subsection{The Stress Tensor Equations}

With our model for the interparticle collision term, the steady state
equations for the stress tensor become \beq p_{rr} - (1 - \xi_r)
{\sigma^2 \over 3} - 4 \Omega \tau p_{r\phi} = H_{rr}\label{ren},\eeq
\beq p_{\phi \phi} - (1 - \xi_\phi) {\sigma^2 \over 3} + {\kappa^2
\tau \over \Omega} p_{r\phi} = H_{\phi \phi},\label{rphien}\eeq \beq
p_{zz} - (1 - \xi_z) {\sigma^2 \over 3} = H_{zz},\label{zen} \eeq and
\beq {\kappa^2 \tau \over 2 \Omega} p_{rr} - 2 \Omega \tau p_{\phi
\phi} + p_{r \phi} = 0, \label{rphi}\eeq where $H_{ij} \equiv \tau
\dot H_{ij}$, i.e., the net stirring over the mean free time.

An ``energy equation'' can be derived by summing equations
(\ref{ren})--(\ref{zen}): \beq H \equiv H_{rr} + H_{\phi \phi} +
H_{zz} = \xi \sigma^2 + 2 p_{r \phi} \Omega' \tau, \label{energy} \eeq
where $\Omega' = d \Omega/ d \ln r$ and $3 \xi = \xi_r + \xi_\phi +
\xi_z$.  For $\Omega' < 0$, equation (\ref{energy}) shows that, if
there is no external stirring, transport must be outwards or zero in
steady state.  This is because any energy lost due to inelasticity
must be extracted from the shear in the mean f{low, which requires
$p_{r \phi} > 0$.  No such thermodynamic constraint on the direction
of angular momentum transport exists if there is external stirring.

\subsection{Transport in a Keplerian Disk}
\label{kepler}

The general solution to equations (\ref{ren})--(\ref{rphi}) is given
in the Appendix.  Here we focus on the case of a Keplerian disk, for
which $\kappa = \Omega$; in the next subsection we briefly contrast
the Keplerian disk with a constant angular momentum disk.  We define
$x = \Omega \tau$ and solve for the $r \phi$ component of the stress
tensor: \beq p_{r \phi} = {x \sigma^2 \over 2} \left({ {4 \over 3}
\left[1 - \xi_\phi \right] - {1 \over 3} \left[1 - \xi_r\right] -
{\H_{rr}} \xi + 4 {\H_{\phi \phi} } \xi \over 1 +
x^2\left[2.5{\H_{rr}} + 10 {\H_{\phi \phi}} + 4
{\H_{zz}}\right]}\right), \label{torque} \eeq where $\tilde H_{ii} =
H_{ii}/H$.  The velocity dispersion of the disk is not an independent
parameter but is determined by solving the equation \beq \sigma^2 = {H
+ x^2 \left(2.5 H_{rr} + 10 H_{\phi \phi} + 4 H_{zz}\right) \over (1 +
4 x^2) \xi - 0.5 \xi_r x^2 + 2 \xi_\phi x^2 - 1.5 x^2}. \label{disp}
\eeq

\subsubsection{Real Particle Disks:  Isotropic Inelasticity}

For a real particle disk with isotropic inelasticity ($\xi_r =
\xi_\phi = \xi_z = \xi$) equations (\ref{torque}) and (\ref{disp})
become \beq p_{r \phi} = {x \sigma^2 \over 2} \left({ 1 - \xi -
{\H_{rr}} \xi + 4 {\H_{\phi \phi} } \xi \over 1 +
x^2\left[2.5{\H_{rr}} + 10 {\H_{\phi \phi}} + 4
{\H_{zz}}\right]}\right) \label{torquepart} \eeq and \beq \sigma^2 = {H +
x^2 \left(2.5 H_{rr} + 10 H_{\phi \phi} + 4 H_{zz}\right) \over (1 +
5.5 x^2)\xi - 1.5 x^2}. \label{disppart} \eeq

In the limit of no external stirring, we recover the usual results of
particle disk theory.  In such an isolated disk, the inelasticity
adjusts to the optical depth of the disk, \beq \xi = {1.5x^2 \over 1 +
5.5 x^2}. \label{xi}\eeq This follows from equation (\ref{disppart})
by setting $H = 0$ and requiring finite $\sigma^2$.  Since the
collisional inelasticity depends on the relative impact velocity,
equation (\ref{xi}) determines a velocity dispersion (Goldreich \&
Tremaine 1978); call this $\sigma_0$.  Inserting equation (\ref{xi})
for $\xi$ in equation (\ref{torquepart}), we confirm that the
transport is always outwards, with\footnote{To derive
eq. [\ref{torquenoH}], write eq. [\ref{torquepart}] as $p_{r \phi} =
0.5 x \sigma^2 N/D$.  Rewritting the denominator as $D = 1 + 4 x^2 +
x^2(6\H_{\phi \phi} - 1.5 \H_{r r})$ and substituting eq. [\ref{xi}]
for $\xi$ into the numerator, $N$, all terms $\propto H_{ij}$ cancel,
yielding eq. [\ref{torquenoH}].}\beq p_{r \phi} = { \xi \sigma^2 \over
3x} = {x \sigma^2 \over 2 + 11 x^2}. \label{torquenoH}\eeq

The fact that isolated fluid disks appear to be hydrodynamically
stable can be explained mechanistically in the sense that they do not
satisfy equation (\ref{xi}).  For any $x$, equation (\ref{xi}) has a
solution only if $\xi \ \epsilon \ [0,0.27]$.  Simulations suggest
that the work done by fluctuating pressure forces is far less
efficient than this. Therefore, for $H = 0$, the only solution to
equation (\ref{disppart}) is $\sigma_0^2 = 0$, i.e., no turbulence.

What happens if the disk is externally stirred?  For $H \ll \xi
\sigma_0^2$ the disk structure is only weakly affected by the
stirring; corrections to $p_{r \phi}$, e.g., are $O[H/\xi \sigma^2_0]
\ll 1$.  Physically, this is because, for $\sigma_0 \ne 0$, heating
must become significant relative to the ``natural'' velocity
dispersion in order to modify the disk structure.  For a
hydrodynamically stable fluid disk with $\sigma_0 = 0$, however, the
``natural'' velocity dispersion vanishes; thus, for any $H$, the
properties of the external stirring (together with the ``collisions'')
determine the angular momentum transport properties of the disk.

In the presence of significant stirring, i.e., $H \gsim \sigma^2_0$,
the direction of angular momentum transport can reverse; in
particular, the sign of the angular momentum transport is then
determined by the sign of the numerator in equation
(\ref{torquepart}).  Note that the sign of $p_{r\phi}$ does not
explicitly depend on $x = \tau \Omega$.  This is consistent with the
arguments given in \S2 and \S3; only the properties of the collisions
and the stirring determine the direction of angular momentum
transport, not the degree to which the disk is collisional.

From the numerator in equation (\ref{torquepart}) it follows that (1)
energy input along $r$ tends to lead to inward transport ($H_{rr}$
enters with a negative sign), (2) energy input along $\phi$ tends to
lead to outward transport ($H_{\phi\phi}$ enters with a positive
sign), and (3) elastic scattering tends to lead to outward transport
($p_{r \phi} > 0$ if $\xi \ll 1$).  All of these results are readily
understood using the theorem of \S3.  The principal long axis of the
velocity ellipsoid generally has a much larger projection along the
radial direction than the azimuthal direction.\footnote{In the
collisionless unstirred limit, equation (\ref{tdelta}) states that
$p_{rr}/p_{\phi \phi} = 2/(2-S) = 4$ for steady-state Keplerian
disks.} Thus $r$ stirring corresponds to elongating the long axis of
the velocity ellipsoid and so should (and does) lead to inward
transport (and vice-versa for $\phi$ stirring).  The transport only
respects the direction of external stirring, however, if the
collisions are fairly inelastic.  In the elastic limit, collisions
efficiently transfer energy from one direction to another,
isotropizing the distribution function; transport is then always
outwards.

Two special cases of equation (\ref{torquepart}) are of particular
interest:
\begin{enumerate} 

\item{$R$ stirring only: This is relevant for interpreting the work of
Stone, Pringle, \& Begelman (1999) and Igumenshchev \& Abramowicz
(1999) who carried out simulations of advection-dominated accretion
flows, which are radially convective.  In these flows, one can treat
the convection as ``externally'' driven by the viscous dissipation
associated with magnetic fields.  As discussed by Narayan,
Igumenshchev, \& Abramowicz (2000) and Quataert \& Gruzinov (2000),
the convective angular momentum transport is inwards in these
simulations, with $\alpha$ as large as $\approx -0.1$ ($\alpha \approx
p_{r \phi}/\sigma^2$ is the dimensionless viscosity).  As noted above,
it is particularly natural for radial convection to produce inward
transport since radial convection tends to directly expand the long
axis of the velocity ellipsoid.  More concretely, equation
(\ref{torquepart}) predicts that the transport should be inwards in
the presence of $r$ stirring so long as $\xi > 1/2$.  Moreover, for
$\xi \sim 1$ and $x \sim 1$,\footnote{This means that the eddy
turnover time is comparable to the rotational period of the disk, as
is indeed the case.}  $\alpha$ should be $\sim -0.1$ in the presence
of $r$ stirring, consistent with simulations.}

\item{$Z$ stirring only: This is relevant for interpreting the
simulations of Stone \& Balbus (1996) in which vertical convection in
an otherwise stable Keplerian disk generated low levels of inward
angular momentum transport ($\alpha \sim -10^{-4}$).  For a particle
disk, equation (\ref{torquepart}) shows that transport is always
outwards in the presence of $z$ stirring: $p_{r \phi} \propto (1 -
\xi)$.  This can be understood by noting that the $z$ dynamics in thin
reflection-symmetric disks is largely decoupled from the $r$ and
$\phi$ dynamics which determine the direction of transport.  In
particular, energy input in the $z$ direction is transferred to the
$r$ and $\phi$ directions by particle collisions, which attempt to
isotropize the distribution function.  By the theorem of \S3 the
transport must then be outwards.}

\end{enumerate} 

\subsubsection{``Anisotropic Inelasticity'':  Inward Transport with Vertical Stirring}

The analysis of the previous section applies to real particle disks in
which the inelasticity is isotropic.  Interestingly, it shows that
vertically stirred particle disks would transport angular momentum
outwards, opposite to what is found in simulations of hydrodynamic
turbulence.  As noted in \S 4.2, the fluid analogue of inelasticity is
inefficient fluctuating pressure forces; such forces are macroscopic
in origin and therefore need not be the same in the $r$, $\phi$, and
$z$ directions.  We now discuss how such ``anisotropic inelasticity''
can produce inward transport in the presence of vertical stirring.

In the presence of vertical stirring only, equation (\ref{torque})
reduces to \beq p_{r \phi} = {x \sigma^2 \over 2} \left({ {4 \over 3}
\left[1 - \xi_\phi \right] - {1 \over 3} \left[1 - \xi_r\right] \over
1 + 4 x^2}\right) \label{torquez}. \eeq From the numerator in equation
(\ref{torquez}) it follows that elasticity in the $r$ direction tends
to lead to inward transport ($1-\xi_r$ has a negative prefactor), while
elasticity in the $\phi$ direction tends to lead to outward transport
($1-\xi_\phi$ has a positive prefactor).

Inward transport in the presence of $z$ stirring requires $\xi_\phi >
0.75 + 0.25 \xi_r$.  Physically, this means that fluctuating pressure
forces are particularly inefficient at doing work in the $\phi$
direction (cf. Stone \& Balbus 1996).  In fact, in the absence of
explicit $\phi$ stirring, equation (\ref{torque}) shows that the
transport is always inwards if $\xi_\phi \rightarrow 1$.\footnote{This
is analogous to Stone \& Balbus's argument that axisymmetric
simulations of hydrodynamic turbulence will always find inward
transport.}  In order to reproduce $\alpha \sim -10^{-4}$ in the
presence of $z$ stirring, as found in the simulations, equation
(\ref{torquez}) requires (taking $x \sim 1$) $1 - \xi_r \sim 10^{-3}$,
and $1 - \xi_\phi \ll 10^{-3}$.

\subsection{Transport in a Constant Angular Momentum Disk}

Balbus and collaborators have emphasized the interesting point that a
constant angular momentum, $\kappa = 0$, disk is analogous to a linear
shear flow (e.g., Balbus, Hawley, \& Stone 1996).  Indeed, numerical
simulations show that the $\kappa = 0$ disk is unstable to nonlinear
perturbations and produces outwards angular momentum transport.  This
feature is also readily captured by our model.  By equation
(\ref{torque2}) the $r \phi$ stress in a $\kappa = 0$ disk is given by
\beq p_{r \phi} = 2 x \sigma^2 \left({{1 - \xi_\phi \over 3} + \xi
\H_{\phi \phi} \over 1 + 8 x^2 \H_{\phi \phi}}\right), \eeq which is
positive definite. For unstirred $\kappa = 0$ disks in which the
particle inelasticity is isotropic, \beq \xi = {8 x^2 \over 3 + 8 x^2}
\, .
\label{xi2} \eeq In contrast to equation (\ref{xi}) for a
Keplerian disk, equation (\ref{xi2}) can be satisfied for some $x$ for
any $\xi \ \epsilon \ [0,1]$.  For a fluid disk this means that,
regardless of the efficiency of fluctuating pressure forces, outward
transport can be sustained in an unstirred disk.  This is consistent
with the fact that shear flows are nonlinearly unstable.

\section{Discussion}

This paper explores the angular momentum transport properties of
axisymmetric, externally stirred particle disks as a means of better
understanding hydrodynamic turbulence in fluid disks. We are not,
however, advocating that turbulence is a sort of macroscopic kinetic
theory, with colliding turbulent ``blobs'' literally replacing
colliding particles.  Rather, we argue that it is physically revealing
to exploit the similar mathematics governing turbulent fluid disks and
particle disks.

There are one-to-one correspondences between terms in the Boltzmann stress
tensor equation governing particle disks and terms in the Reynolds stress
tensor equation governing fluid disks (\S2). This analogy reveals that
the role of differential rotation and epicyclic motion is the same in
both fluid and particle disks; both the rotation rate and the
epicyclic frequency couple to the components of the stress tensor
identically in the two cases.  Consequently, the transport properties
of fluid disks can differ from those of particle disks only to the
extent that the ``scattering function'' for fluid elements differs
from that of particles.  The formal, and very useful, analogy is that
interparticle collisions in particle disks play the same role as
viscous dissipation and fluctuating pressure forces in turbulent
disks.  Both transfer energy in random motions associated with one
direction to those associated with another, and convert kinetic energy
into heat.

Motivated by the above considerations, we solve for the structure of a
particle disk using a simple model for the collision term in the
Boltzmann equation (\S4).  We allow different collisional
inelasticities along the $r$, $\phi$, and $z$ directions.  Although an
anisotropic inelasticity is not realistic for actual particle
collisions, its introduction serves to make the analogy with fluid
turbulence complete.  The fluid analogue of particle inelasticity is
the inefficacy with which fluctuating pressure forces do work, and
there is no {\it a priori} reason for the latter to be isotropic.  A
final ingredient of our model is the insertion of an external stirring
mechanism, modeled as constant source terms in the $r$, $\phi$, and
$z$ energy equations.  External stirring is relevant for interpreting
numerical simulations in which either vertical or radial convection is
present (e.g., Stone \& Balbus 1996; Stone, Pringle, \& Begelman
1999).

We believe that this particle model captures the transport properties
of fluid disks.  With a single equation for the stress tensor
(eq. [\ref{torque}]) we can reproduce results from standard particle
disk theory and from a number of numerical simulations of hydrodynamic
turbulence.

In the limit of no external stirring, differential rotation is the
only source of free energy in our disks.  The angular momentum
transport induced by turbulence or particle collisions must then be
outwards or zero in steady state.  This is because both particle
collisions and turbulence generate heat, which can only be compensated
for by a positive $r \phi$ stress (see eq. [\ref{energy}]).  To have
outward transport in a Keplerian disk requires that ``collisions'' be
relatively efficient at isotropizing the distribution function (see
eq. [\ref{xi}] and the associated discussion).  This tendency towards
isotropization is normally satisfied in axisymmetric particle disks
such as planetary rings; these systems therefore transport momentum
outwards while maintaining some degree of random motion (turbulence).
By contrast, the hydrodynamic stability of isolated fluid disks can be
understood by noting that, according to simulations, fluctuating
pressure forces are extremely inefficient at isotropization.  Thus
there can be no self-sustaining turbulence in these
systems.\footnote{Kato \& Yoshizawa (1997) have argued that if
pressure fluctuations in turbulent Keplerian disks behave like they do
in shear flows and other turbulent laboratory fluids, the
isotropization is sufficient to yield self-sustaining turbulence.
This assumes, however, that Reynold's stress closure models developed
for laboratory problems are applicable to Keplerian disks.}

In the presence of significant external stirring, there are no thermodynamic
constraints on the direction of angular momentum transport.  Instead,
the sign of the transport is determined by the nature of the stirring
and the properties of the collision term in the Boltzmann equation (or
its suitably defined analogue in the fluid problem); formally, a
necessary and sufficient condition for inward transport is that the
combined tendency of collisions and stirring be to make the long
equatorial axis of the velocity ellipsoid yet longer and the short
equatorial axis yet shorter (\S3).

More physically, for any mechanism of external stirring (e.g.,
convection, spiral shocks, etc.), our model predicts the direction of
transport as a function of the ``inelasticity'' of collisions
(eq. [\ref{torque}]).  The latter is, of course, not known {\it a
priori} for the fluid problem, which underscores the parametrized
approach of our model.  Nonetheless, it is gratifying that, so long as
collisions are suitably inelastic, our model finds inward transport
for both vertical and radial stirring (e.g., convection), with the
transport naturally much more efficient for the latter than for the
former.  Inward transport in the presence of purely vertical stirring
requires, however, an anistropic ``inelasticity,'' for which
``collisions'' are more inelastic in the azimuthal direction than in
the radial direction (\S4.4.2).  By equating inelasticity with
inefficient fluctuating pressure forces, as advocated above, we find
results that are consistent with analytic arguments and simulations of
hydrodynamic turbulence (e.g., Balbus \& Hawley 1998).

To conclude, we note that our analysis expands on several aspects of
the Boltzmann equation model for radial convection in rotating media
considered by Kumar, Narayan, \& Loeb (1995).  These authors treated
``collisions'' between convective eddies using the same scattering
function we have considered (eq. [\ref{scat}]) and found that for
sufficiently inelastic collisions, radial convection would produce
inward angular momentum transport.  Interestingly, we now see that
this inward transport has little to do with the details of
convectively unstable fluids (e.g., the superadiabatic entropy
gradient, correlated temperature and velocity fluctuations,
etc. analyzed by Kumar et al.).  Rather, it is a generic property of
disks in the presence of external ``stirring'' of the radial velocity
dispersion.  Moreover, we have emphasized the analogy between
inelastic collisions in particle disks and inefficient fluctuating
pressure forces in turbulent disks, better establishing the relevance
of kinetic theory models to the fluid problem.

\acknowledgments We thank Steve Balbus, Roger Blandford, Peter
Goldreich, Jeremy Goodman, Pawan Kumar, Ramesh Narayan, Scott
Tremaine, and an anonymous referee for useful discussions and
comments.  EQ is supported by NASA through Chandra Fellowship
PF9-10008, awarded by the Chandra X--ray Center, which is operated by
the Smithsonian Astrophysical Observatory for NASA under contract NAS
8-39073. EC acknowledges partial support from a Caltech Kingsley
Foundation Fellowship.

\newpage
\begin{appendix}
\section{Solution for General $\kappa$}

Defining $x = \Omega \tau$, $y = \kappa \tau$, and $x' = dx/d\ln r =
(y^2 - 4 x^2)/2x$, one can solve equations (\ref{ren})--(\ref{rphi}) to
find:

\beq p_{r \phi} = {x \sigma^2 } \left({ {2 \over 3} \left[1 - \xi_\phi
\right] - {y^2 \over 6 x^2} \left[1 - \xi_r\right] - {\H_{rr}} {\xi
y^2 \over 2 x^2} + 2 {\H_{\phi \phi} } \xi \over 1 + 4 y^2 + {y^2 x'
\over x} {\H_{rr}} - 4 x x' \H_{\phi \phi}}\right) \label{torque2} \eeq
and
\beq \sigma^2 = {H + 4 H y^2 + {y^2 x' \over x} {H_{rr}} - 4 x x'
H_{\phi \phi} \over (1 + 4 y^2) \xi - {x' \over 3} \left[ {y^2 \over x} \left(1-\xi_r \right) - 4 x \left(1-\xi_\phi \right) \right]}. \label{disp2} \eeq


\end{appendix}

\end{document}